\def\year{2017}\relax
\documentclass[letterpaper]{article} 
\usepackage{aaai17}  
\usepackage{times}  
\usepackage{helvet}  
\usepackage{courier}  
\usepackage{url}  
\usepackage{color}  
\usepackage{graphicx}  
\usepackage{amsmath}
\begin{document}
%
\title{Crowdsourcing Paper Screening in Systematic Literature Reviews}
\author{Evgeny Krivosheev$^{1}$, Fabio Casati$^{1,2}$, Valentina Caforio$^1$, Boualem Benatallah$^{3}$  \\
$^1$  University of Trento, Italy\\
$^2$ Tomsk Polytechnic University, Russia\\
$^3$ University of New South Wales, Australia\\
\{firstname\}.\{lastname\}@unitn.it, boualem@cse.unsw.edu.au\\
}
\maketitle
\begin{abstract}
Literature reviews allow scientists to stand on the shoulders of giants, showing promising directions, summarizing progress, and pointing out existing challenges in research. At the same time conducting a systematic literature review is a laborious and consequently expensive process. In the last decade, there have a few studies on crowdsourcing in literature reviews. This paper explores the feasibility of crowdsourcing for facilitating the literature review process in terms of results, time and effort, as well as to identify which crowdsourcing strategies provide the best results based on the budget available. In particular we focus on the screening phase of the literature review process and we contribute and assess methods for identifying the size of tests, labels required per paper, and classification functions as well as methods to split the crowdsourcing process in phases to improve results. Finally, we present our findings based on experiments run on Crowdflower.
\end{abstract}
\section{Introduction}

A Literature review is a form of scientific research (and of  publication) that has a high impact on science and society \cite{Sun16-hcomp}. 
Reviews can take different forms and have different objectives 
\cite{Grant2009ReviewTypes}. The main distinction is between \textit{systematic} approaches, where a specific process is defined before the review starts and is followed throughout the identification and analysis of relevant literature, and \textit{non-systematic} ones, where authors do not follow a predefined method for locating and assembling literature.

Literature reviews, especially when systematic, directly provide scientific results and are at the heart of evidence-based approaches, with a potentially profound impact on society \cite{Haidich2010}.
Reviews are also very helpful in introducing newcomers to challenges and opportunities for research in a given area. Not surprisingly, they are among the most highly cited papers (a  search we conducted over a few thousands papers on Scopus shows that the median number of citations for reviews exceeds the median for papers by over 10 times).

Because of their importance and impact, the number of published reviews is rapidly growing\cite{ModernizingSLR13}. This is particularly true for systematic reviews and meta-analyses, in the past popular mostly in the medical field but now widely adopted in all areas of science. 

However, reviews are very time-consuming and effort-intensive. While there are no published statistics on the entire review process (from idea to publication) we are aware of, a study we are conducting with researchers from different fields points to durations of 6 months to 3 years from initial search to submission\footnote{Published data in the healthcare domain indicate that the median time from the \textit{final} literature search to publication in a systematic review is 61 weeks - with the additional problem that over time the list of candidate papers for inclusion becomes out of date and needs to be refreshed \cite{Sampson2008}. However the paper does not report on lag from initial search.}. 
Review results should also be updated periodically, but again the effort for doing so often represents a barrier \cite{Takwoingi2013}.


In this paper we investigate the possibility of crowdsourcing specific aspects of systematic literature reviews. 
We focus specifically on identifying the in-scope papers after initial literature search, and we investigate if and how this phase can be sourced from the citizens, what are the best strategies for doing so, and what is the resulting quality and cost, both in general and compared with the case where the same phase is done by the research team (typically, the co-authors). 
This is a critical phase of a systematic review: not only is it time-consuming (several people work on it, and the combined person-month effort is of over two months), but it is also where risk of bias lies. 

More specifically, we contribute i) a probabilistic model for reasoning over the problem, for tuning the parameters of crowdsourcing tasks to minimize errors, and for providing review authors with information of budget vs error trade-offs, and ii) a set of crowdsourcing strategies and a set of algorithms that minimize the classification error as we vary the assumptions on the model and the model parameters. 
Both the model and the strategies descend from experiments  run on \textit{Crowdflower}\footnote{www.crowdflower.com} and are mindful of what we can actually achieve with some of the practical constraints of typical crowdsourcing platforms.
Experiments on Crowdflower are also used, in addition to theory and simulation, to validate the results as well as to derive parameters for the typical population of workers for this kind of tasks.

Last but not least, experiments provided many insights on task design, such as how the problem should be framed to increase participation and reduce errors, as well as actual pay scales considered acceptable by the community.

\section{Background and Related Work}

\subsection{PRISMA and Systematic Reviews}
Before discussing methods and results we summarize methods and practices for systematic reviews. 
A systematic review follows a defined process and has transparency and clarity as its focal points throughout the whole procedure \cite{khan2003five}. 
This process usually includes (i) the definition of a research question in a clear, structured and unambiguous way; (ii) the identification of all relevant papers through a search strategy that stems from the research question and specifies inclusion and exclusion criteria; (iii) the critical assessment of the included studies; (iv) the data extraction and synthesis in a standardized form, possibly with statistical analysis (meta-analysis); (v) the interpretation of the findings and exploration of any risk for bias \cite{khan2003five,wright2007write,harris2014write,henderson2010write}.

With the objective of increasing the quality of systematic reviews and meta-analyses, the PRISMA statement (\textit{Preferred Reporting Items for Systematic Reviews and Meta-Analyses}) was devised as a guideline to help authors report their reviews in a clear and consistent way \cite{moher2009preferred}. 
As an evolution from the QUOROM statement \cite{Moher1999improving},  PRISMA consists of a 27-items checklist enumerating the details to report and a flow diagram showing the phases of the selection process. Such statements are often required in any systematic review today and are essential in the medical field, where poorly reported reviews can potentially have an effect on people's health. Indeed, Clinical Practice Guidelines, i.e., ``statements that include recommendations intended to optimize patient care", are ``based on systematic reviews of evidence" and should ``be based on an explicit and transparent process that minimizes distortions, biases, and conflicts of interest" \cite{steinberg2011clinical}. Therefore, omitted details and lack of transparency can make this process difficult and contribute to low-quality, misleading guidelines. 


\subsection{Crowdsourcing and Science}
Crowdsourcing is being increasingly adopted as a tool for supporting research \cite{Law_2017_uncertainty}. There are literally hundreds of \textit{citizen science} projects that leverage crowdsourcing at one phase or another of the research, in all fields of science, from biology to astronomy to human sciences \cite{garneau2014crowdsourcing,swanson2015snapshot,hennon2015cyclone,lintott2008galaxy}.
The interest in citizen science has generated a growing body of research on various aspects of the process, from understanding how researchers perceive it \cite{Riesch-epistemological,Law_2017_uncertainty}, to the motivations behind citizens' participation \cite{dabblers14,Frey_motivation_2001}, as well as process and system design \cite{citsci_platform15}.

While all aspects of citizen science research are somewhat interesting and related to this paper, one item of particular importance is the understanding of the conditions under which researchers are motivated to (or deterred from) adopting a crowdsourcing approach. A beautiful analysis of these aspects is provided in \cite{Law_2017_uncertainty} who underscore that one of the concerns is related to how \textit{reviewers} perceive crowdsourcing in research. In other words, crowdsourcing is viable if i) the authors feel that is feasible and valuable for their specific research problem and ii) the authors perceive that reviewers will find it acceptable. This is relevant because literature reviews go through peer reviews and as such the process needs to be accepted by reviewers - and by the community.

The prior art also includes several ``spot" attempts at adopting some form of crowdsourcing in literature reviews. These papers provided inspiration for us although in many cases they are initial, one-off, and relatively small experiments that do not study the variations of the results of crowdsourcing tasks in terms of content and parameters of the experiments, and that in general do not have sufficient statistical power to derive conclusions and guidelines.


Sun and colleagues (Sun et al. 2016) study the feasibility of crowd users in performing the task of extraction information related to interventions from papers’ abstracts in biomedical domains. This preliminary study inferred that giving more concrete examples in the instruction part can help workers be more aware of the purpose of a task. 
A platform for crowdsourcing narrative literature reviews is proposed by Weiss \cite{Weiss2016}, along with insight about challenges appearing in systematic literature reviews in new domains. 
Nguyen et al. \cite{Nguyen2015} proposed an active learning approach to solving the problem of 
deciding whether or not a paper is relevant to a review. 
The authors tried to achieve maximum performance at minimal cost by intelligently choosing between crowd users and domain experts to minimize the expected loss.
They performed experiments on Amazon's Mechanical Turk to classify papers from 4 datasets. 
For each paper, 5 crowd labels and 1 expert label were collected and then evaluated in terms of quality (i.e. False Positives and False Negatives).
Consensus amongst crowd labels is reached through majority vote, whereas the expert label is considered gold data. 

Ng and colleagues \cite{Ng2014} ran a randomized pilot study aimed at exploring the accuracy of medical students  
to perform citation screening via four different modalities, namely a mobile screening application, paper printed with titles and abstracts, a reference management software and a web-based systematic review platform. 
Students were asked to say whether a list of papers were included or excluded from the scope of a review, based on a list of inclusion criteria. 
In case of insufficient information, participants could set papers as ``unsure". 
Participants had never conducted a review, however they had some level of expertise in the field and had received some training in the development of critical appraisal skills, which differs widely from asking to a non-expert crowd to perform such a task. 

\subsection{Modeling and Optimization}

Extensive research, dating back to the 1700s, has addressed the problem of eliciting reliable labels from a crowd, coping with cheating behavior while keeping costs low \cite{karger2011iterative,whitehill2009whose,smyth1995inferring,karger2011budget,hirth2013analyzing,liu2013scoring,eickhoff2013increasing,hirth2011cost}.

One of the first scientists to study this was the Marquis of Condorcet. Condorcet, in his famous \textit{Jury Theorem}\footnote{http://www.stat.berkeley.edu/~mossel/teach/ SocialChoiceNetworks10/ScribeAug31.pdf} of 1785, discusses the probability of a group of persons taking, collectively, the correct classification decision. He shows that if the probability of a person taking the correct decision is greater than 0.5 and votes are independent of each other, then the probability of taking a correct majority decision grows with the number of participants and approaches 1 at the limit (this is, in fact, a direct consequence of the law of large numbers).

From there, a large body of work starting with \cite{DawidSkene_Confusion} and then on to \cite{whitehill2009whose}, estimating labeling in the presence of items of different difficulties, and \cite{LiuWang_truelabel} who apply EM to labeling in the presence of confusion matrix inspire our approach. We also build on insights from Hirth and colleagues \cite{hirth2013analyzing}, who discuss the problem of cost optimization providing information on which cheating detection and task validation algorithms to choose based on the cost structure of the task. 
Our work differs in that we seek for a method to provide, for each task, review authors with a description of price vs error trade-off, an optimal choice of parameters for a given price, and a set of crowdsourcing strategies that aim at minimizing error estimates.


\section{Model and Assumptions}

\subsection{Task Model}
\begin{figure}[htb]
    \centering
    \includegraphics[width=0.45\textwidth]{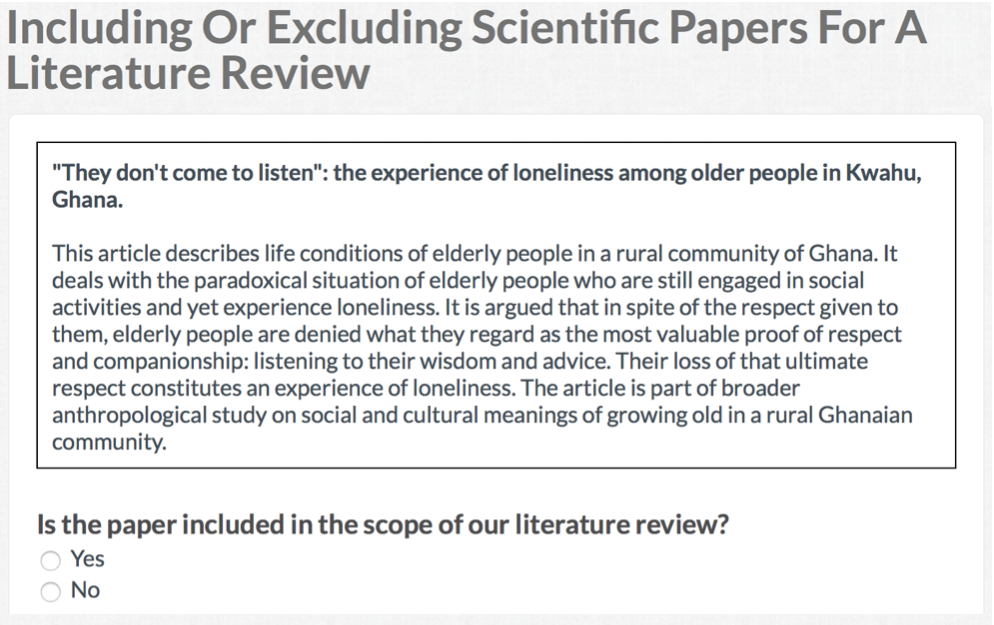}
    \caption{Example of scope-based screening task}
    \label{fig:exp-tasks}
\end{figure}

The crowdsourcing task model includes set of candidate papers $CP=\{p_1,p_2,...p_n\}$ and a textual definition of the scope of the review $S$. The task is performed by workers in a pool of contributors. In practice this pool is very large and for our purposes we assume it is infinite. 
We then ask each worker $j$ to label one or more candidate papers as \textit{in} (the paper is in scope or we do not have sufficient evidence to exclude it from the abstract and title) or \textit{out}, based on $S$. In case of exclusion, they are asked to provide reason to do so. 
Figure \ref{fig:exp-tasks} shows an example task for a review we recently completed. 

The result of the execution of a task is a set of votes $L= \{l_{pj}\}$ representing the binary vote of worker $j$ on paper $p$. Given the set of votes, we use a classification function $cls(L)$ that takes the individual votes and aggregates them to derive the in/out decision for each paper. 

Finally, we define the costs (\textit{loss}) for each error: a cost for a false exclusion $Cost_{fe}$ (deciding that a paper is out when it should have been in), and for a false inclusion $Cost_{fi}$. 
For simplicity we model them as a cost ratio $CR$ that defines how more costly a false exclusion is with respect to a false inclusion. False exclusions are typically more costly since we may be missing an important paper, while a false inclusion ``simply" means that experts will need to go over that paper again. The value of $CR$ depends on the subjective opinion of review authors.

A run of a crowdsourcing task proceeds as follows\footnote{The choice of the model is also guided by what we can do today with platforms such as Crowdflower}: first each worker is shown a set $T$ of test questions with known labels. If the worker answers them correctly, they move to the work phase, where they can provide useful votes (that is, label unknown candidate papers). 
Even during the work phase, test questions are injected with a frequency $tf$ and the contributor is considered trusted (their results are not discarded) only if they keep answering the test questions correctly. The run continues until a given number of labels per paper $J$ has been reached.
In the simplest case a task will have just one run, but we can envision that a run may leave us with uncertainty over some papers and we may want to have additional runs focused on uncertain papers\footnote{In practice, depending on the task settings, it may not always easy to enable a worker to label many papers due to the fact that many concurrent workers access the task in parallel and the available work finishes very quickly.}.
 
Last but not least, each task has a \textit{price}. The price tag is affected by: i) the unit cost per label (how much we pay workers for labeling a paper or an exclusion criterion), ii) the total number of votes asked, and iii) the number of test questions. 
The first two are rather obvious, while the third requires an explanation: with infinite workers, to get accurate results we might simply have a very large number of test questions so that we are sure that only trusted, competent workers remain in the task. In many systems test questions are not paid, so this costs zero. In practice this is not possible: the ethics of this are questionable at best, non-cheaters would do a lot of unpaid work, and we, as task providers, would get bad ratings, impacting our future ability to crowdsource. 
In this paper we take this into account by increasing the price per judgment by a factor $\frac{N_{l}+N_t}{N_{l}}$, where $N_t$ is  the number of initial test questions and $N_{l}$ is the number of valid judgments (i.e., number of votes from a worker who remains above the threshold \textit{tr}) that a worker gives on non-test papers. This essentially states that people who pass the test are in fact paid also for test questions. 
As $N_{l}$ grows our factor becomes ineffective and others can be chosen, but in our case $N_{l}$ is small (as discussed later). Alternative models can be derived, also including a penalty for high test frequencies, but for presenting the concepts and ideas of this paper this is sufficient. The classification cost for a paper is therefore expressed as follows, where $US$ is the cost per vote and $\frac{N_{l}+N_t}{N_{l}}$ is the corrective factor.

\begin{equation}\label{ppp}
PPP = UC \cdot J \cdot \frac{N_{l}+N_t}{N_{l}}
\end{equation}

 

In the end we want to perform candidate paper selection with high accuracy (minimizing the loss) and minimal price. A specific point of interest lies in whether the crowd can achieve an accuracy similar to (or better than) that of experts at a comparable cost, while ensuring transparency and impartiality of the whole process.

\subsection{Probabilistic Model}
To reason about the model and identify strategies and parameters we define a probabilistic model that describes the characteristics of i) tasks and ii) workers. 
Both come into play to identify the optimal crowdsourcing strategy and to set the crowdsourcing parameters.

With respect to the task, we model the following:

\begin{enumerate}
\item Our belief on the proportion of candidate papers that should be included. This is important because it affects the classification function. We do not assume that authors necessarily have such a prior belief, and we discuss later how this parameter can be set or estimated.
\item The \textit{difficulty} level of each paper and of each exclusion criterion: we need to account for the fact that not all candidate papers and exclusion criteria are equal, meaning that some papers may be harder than others to classify, and some criteria may be harder than others to understand. In this paper we  model this parameters with a uniform distribution (which we can parametrize with a variety of priors, such as the commonly used $Beta(\alpha,\beta)$ or priors as suggested in \cite{whitehill2009whose}).
\end{enumerate}

With respect to the workers, we assume that in the worst case workers answer randomly, which means a 0.5 probability of a correct label. The proportion of cheaters is modeled by a Bernoulli random variable $Z$. For non-cheaters, we initially assume a uniform accuracy from 0.5 to 1. 
The accuracy probability function is therefore a mixture of a point mass at 0.5 and a density in the (0.5,1) range.

\begin{equation}
pdf(a) = z \cdot \delta(a-0.5)+ 2 \cdot (1-z) \\
\end{equation}
for $0.5 \leq a < 1$. 
In the function, $\delta$ is the impulse function, while the uniform density is multiplied by $2$ (as it is in the $(0.5,1)$ interval only) and by $(1-z)$ as the density applies only to non-cheaters.
In this paper we do not include more complex cases that include a confusion matrix or priors on the initial probability, but the concepts can be extended to that case. 

\section{Calculating error cost and price}
Now that we have a model we can reason about \textit{strategies} for crowdsourcing literature reviews and assess them based on \textit{assumptions} we can make related to the model.

The  goal is  to i) identify which aspects of the model impact the selection of strategies and results, ii) estimate the model parameters (or at least refine our prior, when available) based on actual experimental data, and iii) derive which strategies can lead to good results in terms of error cost (loss) and price.
Because each problem is different (and even varies also depending on how we title or present the task, as discussed later), the statistical parameters will also vary, so while we can inform our priors via experiments, each task may have to refine the estimation on the go. 

We begin by studying a simple version of the model and a simple crowdsourcing \textit{strategy}.
In general, a crowdsourcing task for literature review can be comprised of a number of \textit{runs}, where in each run $k$ we submit a subset $CP_k$ of the candidate papers $CP$ to the crowd, collecting a given number $J^k$ of labels per paper. 
Furthermore, we start each run with a belief $B^k$ on the proportion of papers to be included, if available (and initially assumed to be 0.5 if there is no estimate). \textbf{XXX check: in figure 2, the theta=0.5 is how the data is, or is it our initial belief assumption or both?}
A run $R^k$ is therefore defined by a tuple $(CP^k, T^k, J^k, TF^k, B^k)$. 

In the simplest strategy the task consists of one run where we submit all papers and seek for $J$ votes per paper.
A classification function will then classify the paper based on the cost ratio $CR$, trying to minimize the loss while fitting within an experiment budget.



The objective for the algorithm here, before even proceeding with the classification, is to i) estimate the optimal values for task parameters that we (as managers of the crowdsourcing process) can play with, such as number of test questions $N_t$, the requested judgments per paper $J$, and the classification function, and ii) provide the scientists with a \textit{budget vs expected loss} curve, showing the error cost depending on the budget, assuming that for each budget we choose the best (lowest loss) configuration identified. The only input explicitly required by the authors is the cost ratio, which is subjective.

The expected error cost (loss) for each paper is given by formula \ref{eq:loss}, where P(FE) and P(FI) denote the probability of false exclusion and false inclusion. 

\begin{equation} \label{eq:loss}
Loss = cr \cdot P(FE) +P(FI)
\end{equation}

Considering that we obtain $J$ judgments per paper, if we decide to exclude a paper after we obtain $J_t$ exclusion votes or more for such a paper, the probability of a false exclusion is given by equation \ref{eq:fe}, where  $\theta_i$ is the (initially unknown) probability that the correct decision for a paper is inclusion, and $\overline{a}_s$ represent the expected accuracy of workers who pass the test phase.
The formulas descend from the observation that  
P(FE) =  $P(decision=exclude / correct=include) \cdot P(correct =include)$, and vice versa for P(FI).

\begin{equation}\label{eq:fe}
P(FE) = \theta_i \cdot  \sum\limits_{{J_t}\leq k \leq J} \binom{J}{k} \cdot (1-\overline{a}_s)^{k}\cdot \overline{a}_s^{J-k}\\
\end{equation}

\begin{equation} \label{eq:fi}
P(FI) = (1-\theta_i) \cdot  \sum\limits_{J-J_t < k \leq J} \binom{J}{k}  (1-\overline{a}_s)^{k}\overline{a}_s^{J-k}
\end{equation}

In this formula, $\theta_i$ is an unknown parameter we need to estimate, $\overline{a}_s$ is also an unknown parameter on which, however, we can have some control by adjusting the test questions $N_t$ to filter inaccurate workers, while $J$ and $J_t$ can be set to optimize loss.



\textbf{xxx to consider renaming either Nt or Jt, as the "t" has different meanings in them}

The accuracy $\overline{a}_s$ of the population that survives $N_t$ tests is distributed as follows: if we denote with $z_s$ the proportion of cheaters in the population that survives the test, which can be derived from Bayes ($z_s= P(test\_passed/cheater)*P(cheater) / P(test\_passed)$), then

\begin{equation} \label{eq:avg}
Z_s = \frac{z \cdot 0.5^{N_t} }
{(z \cdot 0.5^{N_t}) + (1-z)\frac{2}{N_t +1} \cdot (1 - \frac{1}{2^{{N_t}}+1}) }
\end{equation}

Consequently, by using again Bayes for deriving how the accuracy of non cheaters, initially uniform, is reshaped by  the test questions, we obtain:
\begin{equation}
f^t(a) =z_s \cdot \delta(a-0.5) + (1-z_s) \frac{2^{(N_t+1)} \cdot (N_t+1)}{2^{N_t+1}-1}  \cdot a^{N_t}  
\end{equation}
for  $0.5 \leq a < 1$



The expected accuracy $\overline{a}_s$ of this population is $E[x] = \int_{0.5}^1 x \cdot f^t(x)dx$ and is therefore shown in Equation \ref{eq:av}. 
\begin{equation} \label{eq:av}
\overline{a}_s = \frac{z_s}{2} + (1-z_s) \frac{2^{(N_t+1)} \cdot (N_t+1)}{2^{N_t+1}-1} (\frac{1-(0.5)^{N_t+2}}{N_t+2})
\end{equation}


\section{Error minimization and error/price tradeoffs}

We begin our discussion on algorithms by assuming a single-run strategy.

\begin{figure}[htb]
    \centering
    \includegraphics[width=0.49\textwidth]{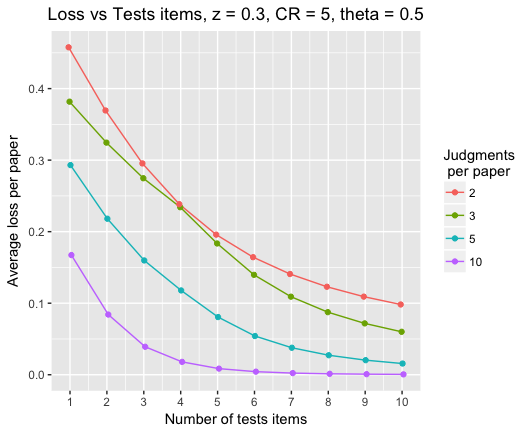}
    \caption{Expected loss depending on the number of test questions and of judgments per paper}
    \label{fig:loss-tests}
\end{figure}

\textbf{Single-run strategy with simple majority voting.} 
In this approach we simply classify papers using majority voting, which is the approach most commonly supported by crowdsourcing platforms. 
For each combination of $N_t$, $N_l$ and $J$ we can compute the total price tag of the experiment as well as estimate the loss via equation \ref{eq:loss}, where $J_t$ is set to $J/2$ (rounded to the upper integer), as shown in Figure~\ref{fig:loss-tests}.
As we have no knowledge of $\theta_i$, we assume a value (such as 0.5, though different values can be set if the task requester has a prior belief). In practice, values of $N_t$ and $J$ over 10 result in near-zero error cost, so computing loss for higher values can be easily done but is rarely needed. 

\begin{figure}[htb]
    \centering
    \includegraphics[width=0.49\textwidth]{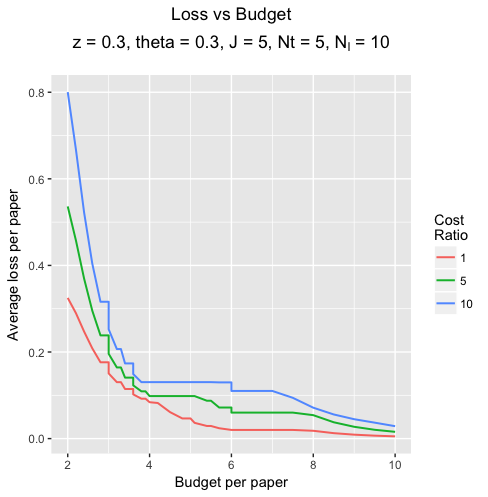}
    \caption{Expected loss that can be achieved depending on the budget}
    \label{fig:loss-budget}
\end{figure}

The result can be plotted as done in Figure \ref{fig:loss-budget}.
The decision of the optimal price/loss point is left to the user as it depends on subjective considerations as well as available budget.
Each budget corresponds to an optimal choice of $N_t$, $N_l$ and $J$ that fits in the budget with minimal loss, so that once we have the requestor's decision we can configure the crowdsourcing task. Notice that for now we are assuming that  our initial guess of $\theta_i$ is correct.


\begin{figure*}[htb]
    \centering
    \includegraphics[width=0.8\textwidth]{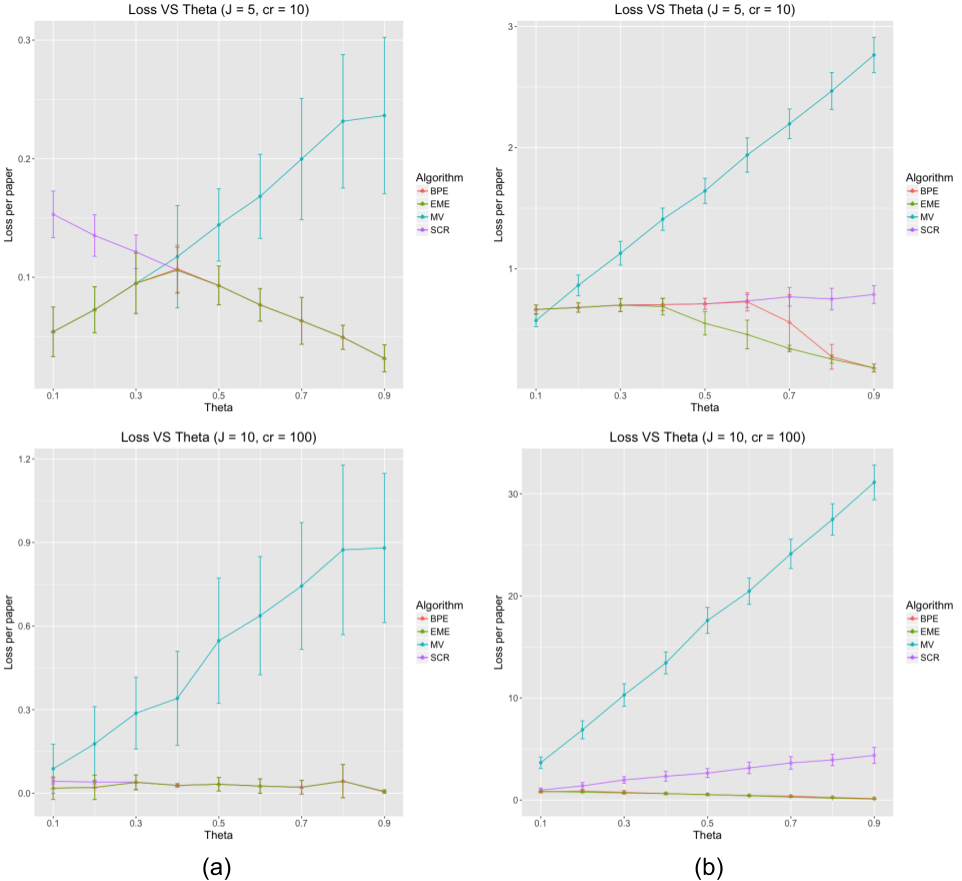}
    \caption{Expected loss for each algorithm. With no difficulty bias (a) and with difficulty bias reducing worker accuracy to a 0.5-0.7 range (b).
    MV is majority voting, SCR is Simple strategy with cost ratio, BPE is basic parameter estimation, and EME is expectation maximization. The simulation is based on 1000 papers}
    \label{fig:lossvtheta}
\end{figure*}

We can then classify the paper using majority voting.
Fig~\ref{fig:lossvtheta} shows the performance of this algorithm (denoted by MV in the legend) in terms of expected loss, assuming an initial run with 5 tests.
Figs~\ref{fig:lossvtheta}(b), top and bottom, differ from Figs~\ref{fig:lossvtheta}(a)
as we assume a more difficult set of papers, in this case simulated by scaling the accuracy of non-cheaters to the 0.5-0.7 range. Furthermore, the top charts have a lower values for $J$ and $cr$. 
The increase with $\theta$ here is due to the fact that this method does not consider the cost ratio which typically penalizes false exclusions. Therefore, error cost grows with the proportion of included papers.

\textbf{Single-run strategy considering cost ratio.} 
The obvious improvement to the baseline is to consider the cost ratio.
This time, for each value of $N_t$ and $J$ we can minimize the loss (according to equation \ref{eq:loss}) by selecting the optimal value for $J_t$.
Again, here we only ``guess" a $\theta_i$ or set it based on the requester belief (in the following chart we assume an initial belief of 0.5).
The minimization can be done using classical minimization algorithms~\cite{arora2015optimization} but also by computing the values given that we have a small number of discrete variables. 
For each combination we have again a price point and we can plot again loss vs price chart, ask the user to point to an acceptable compromise, determine the parameters and run the task as for the previous case.

As we can see from the results in Fig~\ref{fig:lossvtheta}(a) (the label for this algorithm is SCR), this algorithm performs better for high values of $\theta_i$. 
For $\theta_i=0.5$ all algorithms behave similarly as the initial assumption of  $\theta_i=0.5$ holds, while for low value of $\theta_i$ the loss is higher. 
This is because we tend to err on the side of inclusion, so for low values of $\theta_i$ we get higher errors. However for difficult papers where the accuracy is very low, the error actually grows with $\theta_i$, because the probability of false exclusion goes up and if workers are not precise and we do many errors, we pay a price which is not compensated by erring on the side of inclusion. 




\textbf{Single-run strategy with basic parameters estimation.} 
The value of the parameters $\theta_i$ and $z_s$ plays a role in the loss function, and the cost ratio is also important for determining the optimal classification function given the outcome of a run (in our case, for determining $J_t$ which is the only parameter left to play with once we have concluded a run). 
Therefore, we assume we can improve on the above method by estimating $\theta_i$. 
There are many ways in which this can be done. 
One option is to again use \textit{majority voting} but only for performing an initial classification. Based on this, we compute the proportion of included papers and take this as an estimate for $\theta_i$, more informed than an initial guess of 0.5. 
We then compute the accuracy of each worker (as percentage of ``correct" answer based on majority voting classification), and with estimates of $\theta_i$ and $\overline{a}_s$, we then compute the optimal value for $J_t$ based on equation \ref{eq:loss}, and correspondingly we know the minimal loss we can achieve for each price.



As we can see from the results in Fig~\ref{fig:lossvtheta}(a) (the label for this algorithm is BPE), this algorithm performs significantly better than the previous ones for all values of $\theta_i$ except 0.5 (where the guess of the simpler algorithms is correct).

\textbf{Single-run strategy with EM-based parameters estimation.} 
We can improve on the above algorithm by iterating over estimates of the parameters until convergence. A common method for doing so is to leverage Expectation Maximization~\cite{EM_1977,DawidSkene_Confusion}.
In our model, the data is presented as a Bayesian network, where there are two types of variables: 1) the observable votes  provided by workers, and 2) hidden variables, such as  $\theta_i$, the workers' accuracy, and the classification for each paper. Via the EM algorithm we computes the correctness of values given the accuracies of the workers that support it. See~\cite{Pasternack2011} for the details and examples of EM-based for data aggregation. The results shown in Fig~\ref{fig:lossvtheta} indicate that EM is equal to basic estimation and slightly better when accuracy is low.

\textbf{xxx would be good to describe the E and M steps and the starting point}


\textbf{Multi-run Strategies}
The big limitation in all of the previous algorithms is that we run the crowdsourcing task ``in the dark". We ``guess" the value of the parameters and, based on this, set the number of tests and of judgments, leaving the optimization to the post-task analysis phase, when the money has been already spent.
We can improve on this by running a small test-run whose purpose is to obtain initial estimates for $\theta_i$ and $z_s$.
Once we get initial estimates, we can compute and plot again the budget vs loss chart, and based on the estimates and within the confines of the budget, minimize the loss, but this time with the ability to modify $N_t$ and $J$ based on the estimates.
We call this a \textit{horizontal multi-run strategy} as we cut the list of papers horizontally.
The approach assumes that the initial sample of papers is representative of the whole set, and in absence of specific knowledge this means that we randomly reshuffle the papers before selecting the initial $P$ papers for the estimation run.

\begin{figure}[htb]
    \centering
    \includegraphics[width=0.49\textwidth]{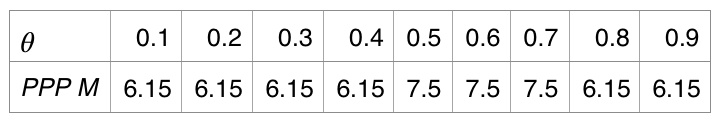}
    \caption{Price per paper with a multirun strategy that has the same loss of a single run strategy at a price of 7.5, optimized with EM-based algorithm. Shown for different values of $theta_i$. Estimation based on $N_t=5, cr=10, z=0.3$, run of 1000 papers.}
    \label{fig:multirun}
\end{figure}

The results are shown in Figure \ref{fig:multirun}, depicting the price per paper we can obtain with a multirun strategy that has the same loss of a single run strategy, run with a budget of 7.5, and optimized with EM-based algorithm. We can see that for values of $\theta_i$ close to 0 or 1 a multirun strategy obtains savings of approximately 20\%.


Multi-run strategies are particularly important when the difficulty of the task is unclear: the difficulty affects the accuracy as pointed out in \cite{whitehill2009whose}, so that papers in certain areas may get lower accuracy than others.
Similarly, we can apply a \textit{vertical} multi-run strategy where we collect \textit{one} vote on \textit{all} papers, and use this to estimate the parameters, and proceed with collecting a second vote, and so on. We omit the details of this for lack of space, but the idea and methods are similar to the horizontal case.

\section{Analysis via Crowdsourcing Experiments}

In the winter and spring of 2017 we run a series of experiments on Crowdflower to assess our results and estimate parameters based on actual crowdsourcing scenarios, as well as to understand how such a task can be framed and how sensitive it is to how we word the question or to the difficulty of the papers. 

We ran a total of 16 experiments with different settings, asking workers to label a total of 50 papers taken from  two systematic reviews, one done by us in an area across computer science and social sciences reviewing technology for fighting loneliness (reference omitted for double blind), using fairly common terminology, and the other in medicine \cite{veronese2017weight} having more complex exclusion criteria, with 26 and 24 papers respectively \footnote{The detailed description of all experiments is available at https://tinyurl.com/csexphc}. 
We collected votes by 2896 respondents (807 of which passed the test phases).
The price of each label per paper was also experimented, ranging from 0.22\$ to 0.35\$, which corresponded to approximately 10 to 15 \$/hour. 
The purpose of the run was not so much to use Crowdflower to get the results, but to understand the workers response in terms of accuracy and speed on real tasks.

The first observation is that the price point is considered acceptable by workers. Overall, the job was rated from 3.3 to 4 in a 5-point scale, and we understand from  Crowdflower that this is above average. 
Interestingly, there is a high variance so that sometimes a lower pay resulted in higher rating for two different tasks with the same settings. Classifications based on exclusion criteria generally get higher ratings for the same pay.
On average the tasks attracted one worker every 20 seconds.
Because of the large pool of workers that end up working concurrently, each worker cannot rate a high number of papers simply because we quickly reach the desired number of votes per paper. 

Another observation is that the worker accuracy changes a lot depending on the subject area. The paper in the medical area, which included complex criteria for determining scope or exclusion obtained an average accuracy of 59\%, versus 83\% for the technology paper. 
Interestingly, the accuracy depends on the title we give to the task, probably as titles that convey that the task is complex tend to discourage the casual worker, and we know that workers correctly perceive task complexity~\cite{yang2016modeling}.
If we word tasks properly and the problem is sufficiently simple, then as shown by Equation \ref{eq:avg} the average accuracy after just a few test questions is very high, and classification errors, even using simple majority voting, are low. 
In this case the classification can be very precise and indeed, in our experiments, in half of the cases (4) where we recorded an ``error" from the crowd, the error was on our side meaning that our ``gold" data turned out to be not so gold. 

We can use the accuracy distributions as derived from Crowdflower and feed them to the algorithms described earlier to compute task settings for relatively easy and relatively hard paper classification problems, and estimate loss for, e.g., a maximum budget of 1\$ per paper and a salary of 20cent per answer. 
For the medical domain case, the optimal algorithm produces $N_t$=10 and $J$=2, giving a cost per paper of 80 cents and an expected error loss for $cr=10$ of 0.15 if $\theta_i=0.5$. 
For reviews where the real $\theta_i=0.1$ the loss is 0.08 for a cost of 1\$ per paper (optimal parameters are $N_t=7$ and $J=3$). 
For the ("easier") technology review we can instead reach a loss of 0.11 when the real $\theta_i=0.5$ (cost of 80cents per paper, $N_t=10, J=2$) and for real $\theta_i=0.1$ the loss is 0.08 for a cost of $N_t=6, J=3$, budget of 96 cents per paper. 

Notice that 1\$ per paper is a reasonable figure: in our preliminary survey of over 20 authors of recent literature reviews, respondents reported an average of 1.5 person-months spent in this phase. 
For a typical screening of approximately 1000 papers the price tag is therefore relatively low.

\section{Summary and Limitations}

The analysis indicates that crowdsourcing literature reviews can be done with high precision and costs figures that are reasonable with respect to the costs spent today. 
Different algorithms can be used to identify the parameters of the crowdsourcing task and the best algorithm we identified based on a multi-run strategy significantly outperforms basic EM (with even larger margins when compared with other simpler algorithms).    
The work has several limitations: in this presentation we could only include a few comparisons and discussions. As the model (and real life scenarios) have many parameters, a more in depth discussion and analysis is needed. 
Furthermore, a deeper understanding of the accuracy as derived from Crowdflower is needed as it is affected by many ``little things" (as we experienced almost by chance) such as the wording of the task title and content and the variability as the type of papers changes. 
Algorithms still have room for improvement, for example in terms of finding the optimal number of papers to consider for the initial run of the multi-run strategy. 
Furthermore, we have not discussed and analyzed the impact of clarity and of ongoing tests submitted to workers who pass the test phases.
A detailed comparison with actual errors performed when experts decide inclusion and exclusion is also needed for a comprehensive evaluation of the approach.






\noindent \textbf{Acknowledgement.} This project has received funding from the EU Horizon 2020 research and innovation programme under the Marie Sklodowska-Curie grant agreement No 690962. 

\bibliographystyle{aaai}
\bibliography{crowdsourcing.bib}

\end{document}